\documentclass[aip,adv,reprint]{revtex4-2}

\usepackage{amsmath}
\usepackage{amssymb}
\usepackage{graphicx}
\usepackage[utf8]{inputenc}
\usepackage{etoolbox}
\usepackage{chemformula}		
\usepackage{siunitx}			
\usepackage{subcaption}			
\usepackage{pdfpages}           
\makeatletter
\AtBeginDocument{\let\LS@rot\@undefined}
\makeatother

\usepackage{hyperref}
\hypersetup{
	colorlinks = True,
	linkcolor = blue,
	citecolor = blue,
	urlcolor = black,}

\makeatletter
\def\@email#1#2{%
 \endgroup
 \patchcmd{\titleblock@produce}
  {\frontmatter@RRAPformat}
  {\frontmatter@RRAPformat{\produce@RRAP{*#1\href{mailto:#2}{#2}}}\frontmatter@RRAPformat}
  {}{}
}%
\makeatother

\newcommand{\UNCC}{Department of Physics and Optical Science, The University of North Carolina at Charlotte, Charlotte, North Carolina, 28223, United States}
\newcommand{\NCTU}{Department of Electrophysics, National Yang Ming Chiao Tung University, Hsinchu 30010, Taiwan}
\newcommand{\RCAS}{Research Center for Applied Sciences, Academia Sinica, Taipei 11529, Taiwan}
\newcommand{\EFMS}{Institute of Physics and Center for Emergent Functional Matter Science, National Yang Ming Chiao Tung University, Hsinchu 30010, Taiwan}
\newcommand{\NSRRC}{National Synchrotron Radiation Research Center (NSRRC), Hsinchu 30076, Taiwan}

\newcommand{\labelphantom}[1]{\parbox{0pt}{\phantomsubcaption\label{#1}}} 

\newcommand{\figref}[1]{Fig.\@ \ref{#1}}
\newcommand{\eqnref}[1]{Eq.\@ \hyperref[#1]{(\ref{#1})}}
\newcommand{\eref}[1]{\hyperref[#1]{(\ref{#1})}}
\newcommand{\tabref}[1]{Table \ref{#1}}

\newcommand{\fluence}[1]{#1$\:\mathrm{mJ/cm}^{2}$} 	

\draft 

\begin{document}
\title{Ultrafast Multi-Shot Ablation and Defect Generation in Monolayer Transition Metal Dichalcogenides}

\author{Joel M. Solomon}
\affiliation{\UNCC}

\author{Sabeeh Irfan Ahmad}
\affiliation{\UNCC}

\author{Arpit Dave}
\affiliation{\UNCC}

\author{Li-Syuan Lu}
\affiliation{\NCTU}
\affiliation{\RCAS}

\author{Yu-Chen Wu}
\affiliation{\NCTU}

\author{Wen-Hao Chang}
\affiliation{\NCTU}
\affiliation{\RCAS}

\author{Chih-Wei Luo}
\affiliation{\NCTU}
\affiliation{\EFMS}
\affiliation{\NSRRC}

\author{Tsing-Hua Her}
\email{ther@uncc.edu}
\affiliation{\UNCC}


\begin{abstract}
Transition metal dichalcogenides are known to possess large optical nonlinearities and driving these materials at high intensities is desirable for many applications. Understanding their optical responses under repetitive intense excitation is essential to improve the performance limit of these strong-field devices and to achieve efficient laser patterning. Here, we report the incubation study of monolayer \ch{MoS2} and \ch{WS2} induced by \SI{160}{\femto\second}, \SI{800}{\nano\meter} pulses in air to examine how their ablation threshold scales with the number of admitted laser pulses. Both materials were shown to outperform graphene and most bulk materials; specifically, \ch{MoS2} is as resistant to radiation degradation as the best of the bulk thin films with a record fast saturation. Our modeling provides convincing evidence that the small reduction in threshold and fast saturation of \ch{MoS2} originates in its excellent bonding integrity against radiation-induced softening. Sub-ablation damages, in the forms of vacancies, strain, lattice disorder, and nano-voids, were revealed by transmission electron microscopy, photoluminescence, Raman, and second harmonic generation studies, which were attributed to the observed incubation in 2D materials. For the first time, a sub-ablation damage threshold is identified for monolayer \ch{MoS2} to be 78\% of single-shot ablation threshold, below which \ch{MoS2} remains intact for many laser pulses. Our results firmly establish \ch{MoS2} as a robust material for strong-field devices and for high-throughput laser patterning.
\end{abstract}


\maketitle 

Understanding the optical responses of two-dimensional (2D) materials under strong excitation is important. For example, semiconductor transition metal dichalcogenides (TMDs) have demonstrated strong optical nonlinearities such as second harmonic generation (SHG),\cite{Mannebach2014, Kumar2013} high harmonic generation,\cite{Liu2017} saturable absorption,\cite{Wang2013, Nie2018} and giant two-photon and three-photon absorption.\cite{Li2015, Zhang2015, Zhou2017a} These properties make TMDs an excellent candidate for attosecond photonics, mode locking, optical limiting, and multi-photon detectors. For such applications, performance is limited by the optical damage of 2D materials so understanding the laser-induced damage in any forms of defects under repetitive excitation is critical to improve their performance metrics. Additionally, ultrafast lasers have been demonstrated to selectively remove 2D materials for specific sizes and geometries.\cite{Park2012, Solomon2021} Compared to electron-beam and photolithography which have high costs, complexity, and degraded device performance due to unwanted dopants, contaminates, and polymer residues,\cite{Kollipara2020} ultrafast laser ablation is a promising technique to pattern 2D materials that is \emph{in situ}, resist-free, and maskless. For this application, knowledge of multi-shot ablation is essential to select optimal laser parameters to deterministically remove material. As the onset of ablation and damage are governed by similar physical principles, determining the ablation threshold fluences $F_{th}$  of 2D materials as a function of the number of pulses $N$ admitting on the same spot of the substrate is important.

Such a phenomenon, commonly referred to as incubation, is well known in pulsed laser-induced ablation and damage of bulk materials.\cite{Sun2015} Starting from the single-shot ablation threshold $F_{th}(1)$, the threshold fluence decreases monotonically with $N$ until it approximately saturates at $N_{sat}$. For $N > N_{sat}$, $F_{th}$ varies in a small range of fluence, defined as $F_{th}(\infty)$. If the fluence is below $F_{th}(\infty)$, ablation does not occur for any number of pulses, at least theoretically. The ratio $R\equiv F_{th}(\infty)⁄F_{th}(1)$ is a measure of the degree of incubation: the larger the $R$, the less pronounced the incubation and the more resistant the material is to radiation damage. We note that the incubation study of graphene was previously reported by Wetzel \emph{et al.\@} but their study was purely experimental without any discussion or understanding.\cite{Wetzel2013} On the other hand, femtosecond multi-shot degradation of monolayer \ch{MoS2} was reported by Paradisanos \emph{et al}.\cite{Paradisanos2014} They observed softening of the Raman $A_{1g}$ and $E_{2g}^{1}$ modes at two fluences (25\% and 40\% of $F_{th}(1)$), which they attributed to a decrease of the Mo-S bond density. Incubation in terms of $F_{th}(N)$ and their signatures beyond Raman scattering for monolayer TMDs has not been reported.

In this work, the multi-shot ablation threshold for monolayer \ch{MoS2} and \ch{WS2} is studied and described based on a phenomenological model for the first time. We also carried out transmission electron microscopy (TEM), and second harmonic generation (SHG), photoluminescence (PL), and Raman spectroscopies to reveal morphological and optical property changes in \ch{MoS2} in the sub-ablation damage regime to provide insight into defect formation. The saturation threshold is found to be about 75\% of the single-shot ablation threshold and coincides with the damage threshold as measured by second harmonic generation. This provides direct evidence that incubation begins with laser-induced defects in 2D materials, including vacancies and lattice distortion.

\begin{figure}[t]
\centering
\labelphantom{fig:1a}
\labelphantom{fig:1b}
\labelphantom{fig:1c}
\includegraphics[scale=1.0]{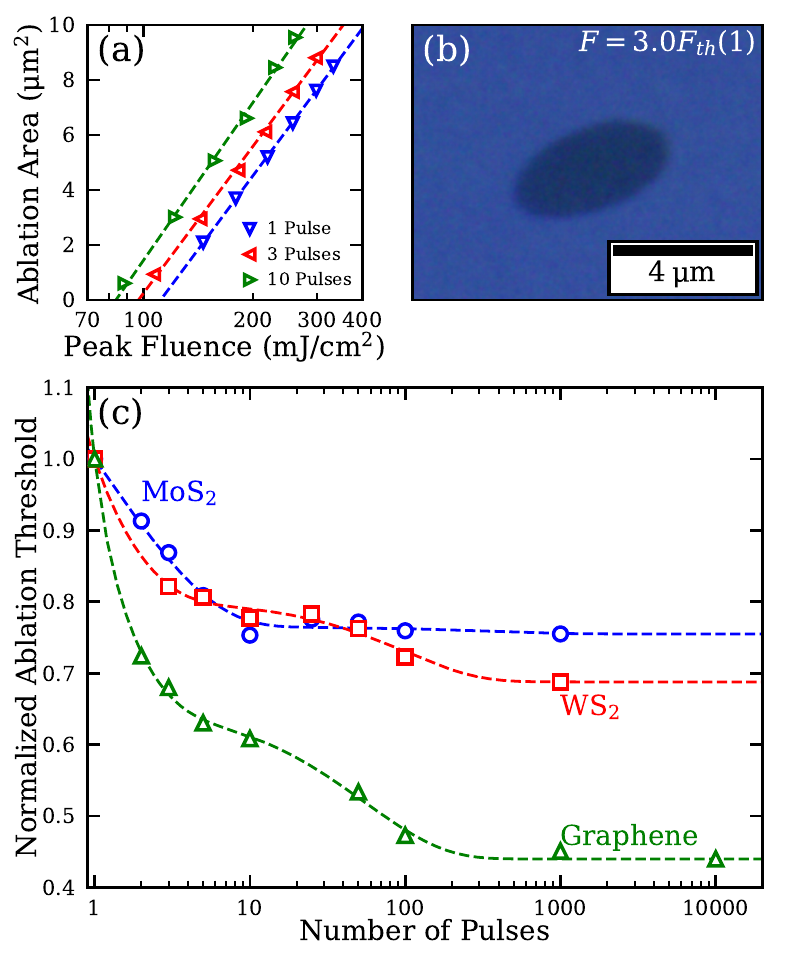}
\caption{(a) Determination of the ablation threshold of \ch{MoS2} for 1 pulse, 3 pulses, and 10 pulses. (b) Optical microscope image of an ablated hole made in a \ch{MoS2} film due to a single pulse. (c) Normalized multi-shot ablation thresholds for monolayer \ch{MoS2}, \ch{WS2}, and graphene.
\label{fig:1}}
\end{figure}

Monolayer \ch{MoS2} films and \ch{WS2} flakes were grown by chemical vapor deposition on \emph{c}-cut sapphire (\ch{Al2O3}) and transferred to a borosilicate glass and \ch{Al2O3} substrate, respectively.\cite{Hsu2019} A Coherent RegA 9000 producing \SI{160}{\femto\second} pulses at a central wavelength of \SI{800}{\nano\meter} is used to conduct all ablation experiments. The repetition rate was kept constant at \SI{307}{\hertz} where a mechanical shutter was used to select a single or multiple pulses from the pulse train. The laser pulse energy was controlled using a continuous neutral density filter wheel or a half-wave plate with a polarizer, where a calibrated, fast photodiode was used to record individual pulse energies $E$. The pulses were focused on to the \ch{MoS2} film by a 10$\times$, 0.26 NA objective which also allowed for \emph{in situ} imaging of the experiment. The \ch{MoS2} sample was mounted on an Aerotech ANT three-axis motorized translation stage for precise positioning control.

The ablation threshold $F_{th}$ was determined by measuring the ablated hole area as a function of the incident peak fluence $F$ where the linear extrapolation of the hole area versus $\ln(F)$ yields the threshold fluence $F_{th}$ at zero area.\cite{Liu1982} This area method is illustrated in \figref{fig:1a} for \ch{MoS2} on a glass substrate and $N=1$, 3, and 10 pulses. For fluences above $F_{th}$, the ablation produces deterministic holes with a well-defined shape, as illustrated in \figref{fig:1b}. The effective laser spot radius was \SI{2.3}{\micro\meter} as measured by both a CCD camera and the fits in \figref{fig:1a}. Figure \ref{fig:1c} shows the normalized $F_{th}$ of \ch{MoS2} and \ch{WS2} for selective $N$ up to 1000 pulses. As a comparison, the incubation for graphene from Ref.\@ \citenum{Wetzel2013} is also shown. For reference, the values of $F_{th}(1)$ and $R$ for each of these materials are summarized in \tabref{tab:1}. For \ch{MoS2}, $F_{th}$ decreases monotonically from $N=1$ to 10 before saturating at $N_{sat}\sim10$. For \ch{WS2} and graphene, $F_{th}$ quickly reduces for the first three pulses and then experiences an inflection, before leveling off at 1000 pulses. The $R$ values for \ch{MoS2} and \ch{WS2} were found to be larger than the majority of bulk materials.\cite{Sun2015, Rosenfeld1999, Mero2005} Only \ch{HfO2} ($R\sim0.73$, $N_{sat}\sim10{,}000$, \SI{50}{\femto\second}) and \ch{Ta2O5} ($R\sim0.67$, $N_{sat}\sim1000$, \SI{150}{\femto\second}) films were found to have comparable $R$ values to \ch{MoS2} and \ch{WS2}, respectively.\cite{Nguyen2008, Mero2005} Among all these materials, \ch{MoS2} has the fastest saturation.

To model the data, we generalize the phenomenological model introduced by Sun \emph{et al.\@} to 2D materials, which was formulated in terms of the change in absorption and critical bulk energy density.\cite{Sun2015} We assume the absorption $A\equiv\Delta E⁄E$ and the critical surface energy density $G^{\prime}$ change over successive pulses according to  
\begin{equation}
\begin{aligned}
A(N,F^{\prime}) &= A_{0} + \Delta A\left(1-e^{-\beta F^{\prime} N}\right), \\
G^{\prime}(N,F^{\prime}) &= G_{0}^{\prime} + \Delta G^{\prime}\left(1-e^{-\gamma F^{\prime} N}\right), 
\end{aligned}
\label{eq:1}
\end{equation}
where $A_{0}$ is the initial absorption, $\Delta A = A_{max}-A_{0}$ is its maximal change (typically positive), $G_{0}^{\prime}$ is the initial critical surface energy density needed for single-shot ablation, $\Delta G^{\prime}=G_{min}^{\prime} - G_{0}^{\prime}$ is its maximal change (typically negative), and the coefficients $\beta$ and $\gamma$ are rate constants. Importantly, $F^{\prime}=\xi F$ is the internal fluence inside the 2D materials where $\xi$ represents the intensity enhancement factor due to the interference effect of the supporting substrate, which are $\xi = 0.63$ and 0.53 for borosilicate glass and \ch{Al2O3}, respectively.\cite{Solomon2021} Using the internal fluence allows the determination of substrate independent coefficients. According to this model, the threshold fluence $F_{th}(N)$ is reached when the energy density deposited by the $N^{\mathrm{th}}$ pulse equals the critical energy density modified by the preceding $N-1$ pulses:  
\begin{equation}
A\left(N-1, F_{th}^{\prime}(N)\right) F_{th}^{\prime}(N) = G^{\prime}\left(N-1, F_{th}^{\prime}(N)\right).
\label{eq:2}
\end{equation}
By combining Eqs.\@ \eref{eq:1} and \eref{eq:2}, the multi-shot threshold satisfies the following transcendental equation
\begin{widetext} 
\begin{equation}
F_{th}^{\prime}(N) = \dfrac{F_{th}^{\prime}(1) - \left[F_{th}^{\prime}(1) - F_{th}^{\prime}(\infty)\left(1+\dfrac{\Delta A}{A_{0}}\right)\right]\left[1-e^{-\gamma F_{th}^{\prime}(N)(N-1)}\right]}{1+\dfrac{\Delta A}{A_{0}}\left[1-e^{-\beta F_{th}^{\prime}(N)(N-1)}\right]}
\label{eq:3}
\end{equation}
\end{widetext}
where $F_{th}^{\prime}(1) = G_{0}^{\prime}/A_{0}$ is the intrinsic single-shot ablation threshold and $F_{th}^{\prime}(\infty)$ is the intrinsic multi-shot ablation threshold at saturation. From \eqnref{eq:3}, $R$ can be expressed as a simple analytical function of the maximal fractional change in critical energy $\Delta G^{\prime}/G_{0}^{\prime}$ and absorption $\Delta A/A_{0}$ according to 
\begin{equation}
R\equiv \dfrac{F_{th}(\infty)}{F_{th}(1)}=\dfrac{1+\Delta G^{\prime}/G_{0}^{\prime}}{1 + \Delta A/A_{0}}.
\label{eq:4}
\end{equation}
To obtain a large $R$ value, \eqnref{eq:4} indicates $\Delta G^{\prime}$ and $\Delta A$ should be as close to zero as possible. Equation \eref{eq:3} is applied to fit experimental data in \figref{fig:1c} to extract the fitting parameters $\Delta A/A$, $\beta$, and $\gamma$. $\Delta G^{\prime}/G_{0}^{\prime}$ is then calculated from \eqnref{eq:4}. The dashed curves in \figref{fig:1c} are the theoretical fits where all the fitting parameters are shown in \tabref{tab:1}. This model provides insight into the incubation behavior of these 2D materials. Though they have similar $\Delta A/A_{0}$ values, $\Delta G^{\prime}/G_{0}^{\prime}$ is negligible for \ch{MoS2}, yielding the largest $R$ and the smallest $N_{sat}$ with a single decay trend in its incubation. For \ch{WS2} and graphene, the initial fast decay of $F_{th}$ is due to a strong saturation of $\Delta A$ with the number of pulses (i.e. large $\beta$), followed by a slow decay due to a weak saturation of $\Delta G^{\prime}$ (i.e. small $\gamma$), leading to a larger $N_{sat}$. The transition between these two decays manifests the inflection in \figref{fig:1c}.    
\begin{table}[ht]
\centering
\caption{Fit parameters for \figref{fig:1c}.
\label{tab:1}}
\resizebox{\columnwidth}{!}{%
\begin{tabular}{r|c|ccccc}
\multicolumn{1}{r}{\textbf{Materials}} & \multicolumn{1}{c}{
\begin{tabular}{c}
\boldmath$F_{th}(1)$ \\
\boldmath$(\mathrm{mJ/cm}^{2})$
\end{tabular}} & \boldmath$R$ & \boldmath$\dfrac{\Delta A}{A_{0}}$ & \boldmath$\dfrac{\Delta G^{\prime}}{G^{\prime}_{0}}$ & 
\begin{tabular}{c}
\boldmath$\beta$ \\
\boldmath$(\mathrm{cm}^{2}/\mathrm{J})$
\end{tabular}  &  
\begin{tabular}{c}
\boldmath$\gamma$ \\
\boldmath$(\mathrm{cm}^{2}/\mathrm{J})$
\end{tabular} \\ 
\hline
\ch{MoS2}	& 111 & ~0.75~ & ~0.31~ & ~-0.013~ & ~6.3~ & ~0.052~  \\
\ch{WS2}	& 161 & ~0.69~ & ~0.25~ & ~-0.14~  & ~13~  & ~0.16~  \\
Graphene	& 156 & ~0.44~ & ~0.55~ & ~-0.32~  & ~15~  & ~0.35~ 
\end{tabular}}
\end{table}

\begin{figure}[ht]
\centering
\includegraphics[scale=1.0]{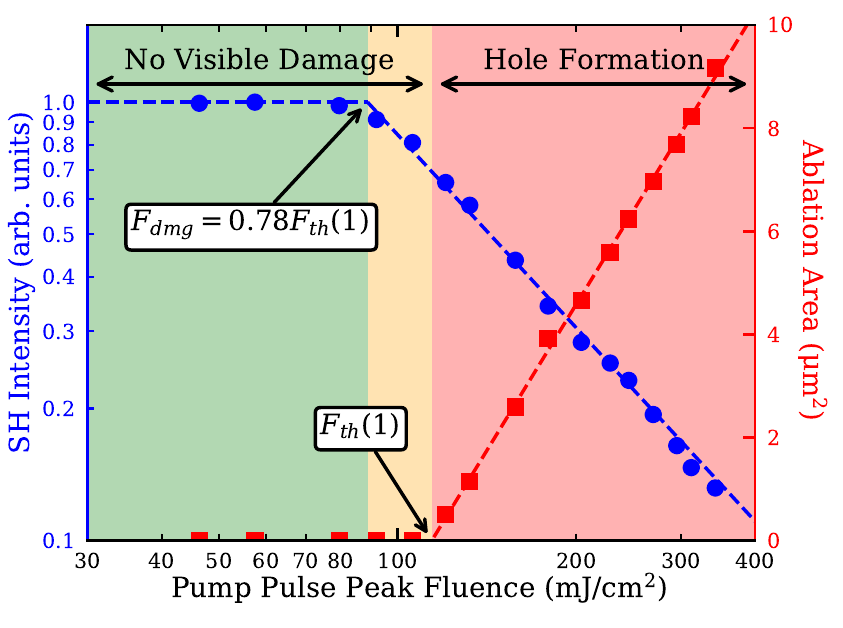}
\caption{SH intensity of damage to the \ch{MoS2} film due to a single pump pulse. The fluences in the red region creates a hole visible under an optical microscope. Fluences in the yellow region damage the film without visible evidence. No permanent damage to the film occurs for fluences highlighted in green. The probe pulse fluence is 30\% of $F_{th}(1)$. The left axis is logarithmic, and the right axis is linear.
\label{fig:2}}
\end{figure}

\begin{figure*}[ht]
\centering
\includegraphics[scale=1.0]{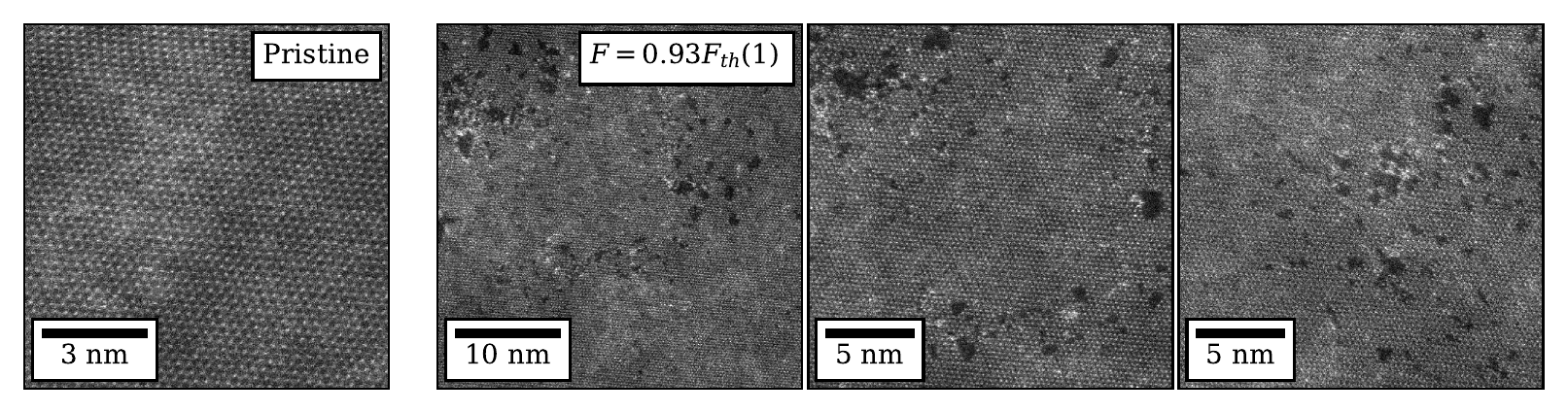}
\caption{(left) HR-TEM of pristine monolayer \ch{MoS2}. (right) Damaged \ch{MoS2} film exposed to a pulse with a fluence $F=0.93F_{th}(1)$.
\label{fig:3}}
\end{figure*}

To gain information on the structural changes induced by the laser exposure, we performed a correlated study of the ablation area and SHG, since SHG is known to be sensitive to changes in crystal structure. In this experiment, the \ch{MoS2} sample is exposed to a single intense pump pulse, followed by a train of weaker pulses to probe the resultant structural modification using static SHG. The result is shown in \figref{fig:2} for various pump fluences below and above $F_{th}(1)$. For fluences above $F_{th}(1)$ where holes are created deterministically, the SH signal decreases with increasing fluence, which can be understood in terms of the reduced spatial overlap of the probe pulse with the edges of the hole. For pump fluences below \fluence{89}, the SH intensity remained the same as that of pristine \ch{MoS2}, indicating the material is intact. Between them where no visible hole was seen, the SH signal is still below the pristine value, indicating that the film is permanently damaged. Accordingly, we define the sub-ablation damage threshold fluence $F_{dmg}=0.78 F_{th}(1)$ as the minimal fluence causing permanent damage to the film. At this limit, the damage most likely consists of localized vacancies and lattice disorder. Importantly, comparing Figs.\@ \ref{fig:1} and \ref{fig:2} yields $F_{th}(\infty)\sim F_{dmg}$. This finding clearly proves that such laser-induced defects represent the beginning stages of incubation for ablation. As long as the pulse fluence is below $F_{dmg}$, monolayer \ch{MoS2} will not be ablated for any number of pulses $N > N_{sat}$. Similarly, for $F_{th}(N)$, each pulse will generate sub-ablation damage, accumulatively creating a deterministic hole at its zero-area limit by the $N^{\mathrm{th}}$ pulse.

To visualize such sub-ablation damage, a \ch{MoS2} film was transferred to a holey carbon film grid for high resolution transmission electron microscopy (HR-TEM). The left image in \figref{fig:3} shows a HR-TEM image of pristine \ch{MoS2} where the dark spots represent the sulfur atoms and the bright spots are the molybdenum atoms. The images on the right show \ch{MoS2} exposed to a single pulse at $F=0.93F_{th}(1)$ where clusters of atoms ranging up to a few nanometers across are removed without destroying the overall integrity of the film. Figure \ref{fig:3} reveals that such sub-ablation damage is stochastic in nature where voids appear sporadic with random sizes and shapes within the exposed area. This is in sharp contrast to the deterministic ablated holes seen in the area method (\figref{fig:1a}). The nano-voids in \figref{fig:3} show a decrease of roughly 5\% in atomic density, as determined by ImageJ analysis, which translates to a 10\% reduction in SH intensity. Figure \ref{fig:2}, however, indicates a 20\% reduction in SH intensity, implying the presence of other defects in addition to the nano-voids. Again, this supports the existence of vacancies and lattice disorder as suggested in \figref{fig:2}, which are beyond the resolution of our TEM.

\begin{figure*}[ht]
\centering
\labelphantom{fig:4a}
\labelphantom{fig:4b}
\labelphantom{fig:4c}
\labelphantom{fig:4d}
\labelphantom{fig:4e}
\labelphantom{fig:4f}
\includegraphics[scale=1.0]{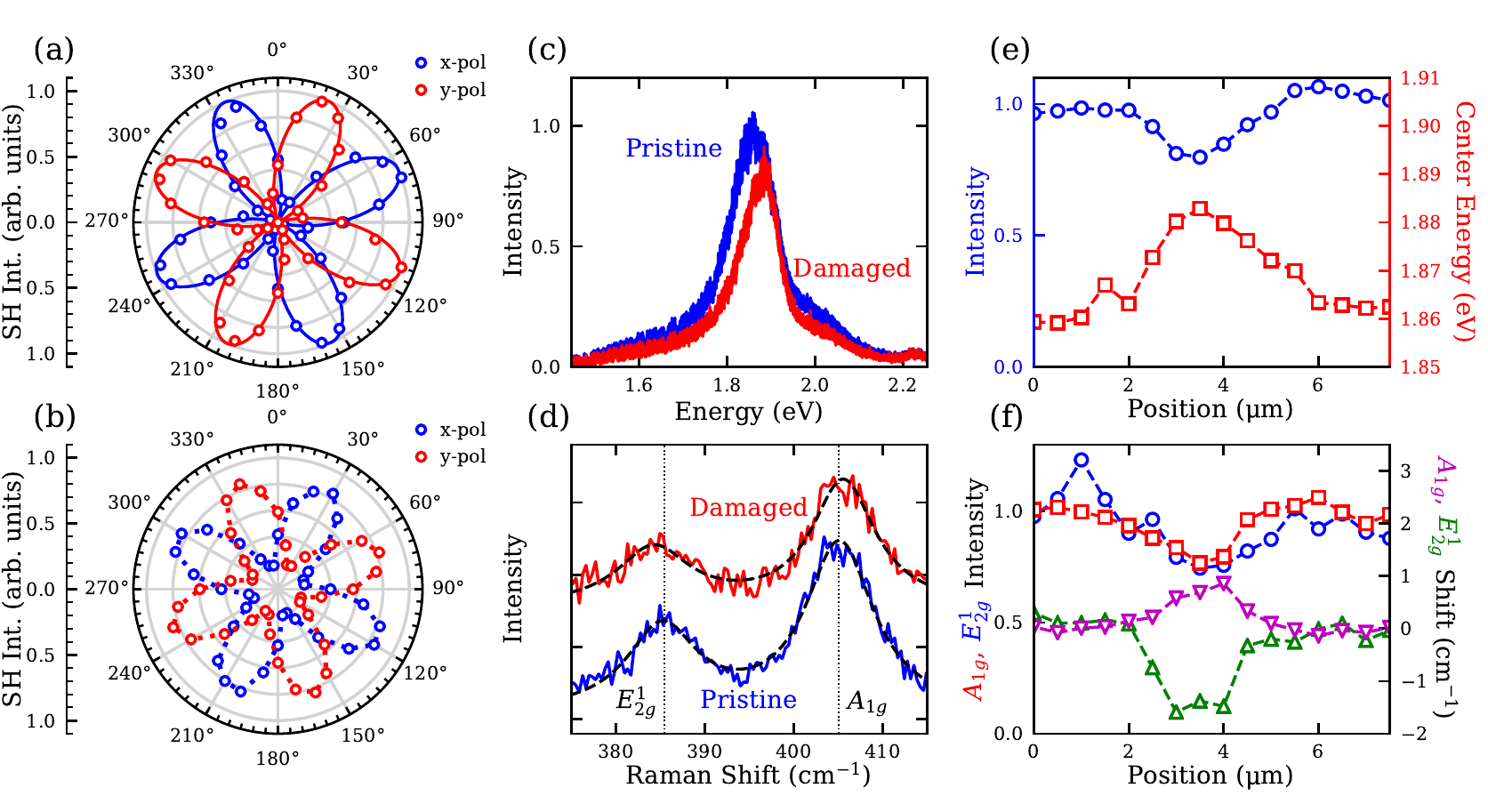}
\caption{Second harmonic polar profile of (a) pristine \ch{MoS2} and (b) damaged \ch{MoS2} by a pulse of $F=0.83F_{th}(1)$. (c) Photoluminescence  and (d) Raman  spectra of pristine and damaged \ch{MoS2}. (e) Photoluminescence and (f) Raman line scans across damaged \ch{MoS2}. The second harmonic polar plot in (b) and the PL and Raman scans in (c)-(f) were all recorded for the same damaged spot. 
\label{fig:4}}
\end{figure*}

To reveal more optical signatures of these defects, we performed various optical spectroscopies on pristine \ch{MoS2} and \ch{MoS2} exposed to a single pulse at $0.83F_{th}(1)$. Figure \ref{fig:4a} shows the SH polar profile for pristine \ch{MoS2} by rotating the incident polarization while recording the SH signal at x- and y-polarizations. The data follows the theoretical curve (solid line) for this method, indicating excellent crystallinity.\cite{Mannebach2014} The polar profile at the center of the damaged region is shown in \figref{fig:4b}. As the damaged region is significantly smaller than the probe pulse spot size, the SH signal is dominated by the surrounding pristine \ch{MoS2} where a 4-fold symmetry is still clearly resolved. Compared to \figref{fig:4a}, however, the depolarization (reduced contrast) is evident where the maximal SH intensity reduces by $\sim$20\% and the minimum never drops completely to zero. A more pronounced depolarization can be seen for \ch{MoS2} exposed to $2F_{th}(1)$ where more of the SH signal is from the edge of the hole (see the supporting information). In addition to SHG, Figs.\@ \ref{fig:4c}-\ref{fig:4d} show the PL and Raman spectra recorded at the center of the damaged spot. As a reference, pristine spectra are also shown in blue. The line profile of the PL peak intensity and center energy scanned across the damaged spot is presented in \figref{fig:4e}, showing an intensity reduction of $\sim$25\% and a blue shift of 0.02 eV after exposure. Similar results were also recorded for the Raman line scan as shown in \figref{fig:4f}, where the intensity of both the $E_{2g}^{1}$ and $A_{1g}$ peaks are reduced by $\sim$25\% in the damage region. Additionally, the $E_{2g}^{1}$ mode experiences a maximal red shift of $1.6\:\mathrm{cm}^{-1}$ while the $A_{1g}$ peak shows a maximal blue shift of $0.9\:\mathrm{cm}^{-1}$.  

For single pulse exposure at $0.83F_{th}(1)$, the reduction of the Mo-S bonds is estimated below 5\% from \figref{fig:3}, indicating other defects play a significant role in the observed 20\%-25\% drop in SH, PL, and Raman intensities. The presence of vacancies and disorder breaks the lattice translational symmetry, which compromises the constructive and destructive interference among SH dipoles across the sample. This will reduce the maximal SH signal and increase the minimal SH signal in the polar plot, consistent with the observed de-polarization effect in \figref{fig:4b}. In addition, dangling bonds associated with these vacancies are known to introduce mid-gap states. Several types of vacancies associated with S and Mo atoms are shown to generate mid-gap states with energies ranging from 0.02 to 1.72 eV within the band gap, which can provide non-radiative decay pathways.\cite{Hong2015, Khan2017, Bahmani2020, Liu2016} These states have been invoked to strongly quench the PL intensity and cause a blue shift similar to that observed here.\cite{Ma2013, Forster2017, Sivaram2019, Oh2016, Lu2015} The blue shift in the PL peak can also be attributed to the adsorption of oxygen on the \ch{MoS2} monolayer given the ambient conditions of the experiment.\cite{Oh2016} Moreover, occupation of these mid-gap states by the preceding pulses could increase the light absorption $(\Delta A/A_{0})$ of successive laser pulses during incubation, leading to a reduction in threshold fluence.\cite{Khan2017} Additionally, the lateral strain introduced locally by these vacancies and lattice disorder can cause the Raman intensities to decrease.\cite{YWang2013, Kukucska2017} The blue shift in the $A_{1g}$ and red shift in the $E_{2g}^{1}$ peaks could be explained by a combination of p-doping from the presence of Mo-O bonds\cite{Mawlong2018, Wu2013} and strain introduced by these defects.\cite{Oh2016, YWang2013}  

In summary, we show that \ch{MoS2} and \ch{WS2} have weak incubation effects when excited by \SI{160}{\femto\second}, \SI{800}{\nano\meter} pulses in air. A modified phenomenological model based on the change in surface energy density and absorption was applied to describe the incubation behavior of \ch{MoS2}, \ch{WS2}, and graphene with good agreement. According to this model, \ch{MoS2} has negligible laser-induced bond softening, yielding its weakest incubation ($R\sim0.75$) and fastest saturation ($N_{sat}\sim10$). For a \ch{MoS2} film exposed to a single pulse at a fluence below $F_{th}(1)$, our TEM analysis reveals nano-voids and our optical spectroscopy data strongly support the presence of atomic-scale defects such as vacancies and lattice distortion, which are responsible for the observed reduction and peak energy shift in PL and Raman, as well as reduced contrast in polarized SHG. Moreover, static SH measurements identify the threshold fluence to generate such defects to be $F_{dmg}=0.78F_{th}(1)$ for \ch{MoS2}, which coincides with $F_{th}(\infty)$ and provides direct evidence that laser-induced defects set the beginning stages of incubation for ablation of \ch{MoS2}. For fluences below $F_{dmg}$, \ch{MoS2} remains intact for an infinite number of pulses. For fluences slightly above $F_{dmg}$, incubation starts with laser generated atomic defects. Successive pulse exposure leads to nano-voids with random sizes and shapes, which grow in size with subsequent pulses, until a deterministic ablation hole with a theoretical zero area is reached. For fluences between $F_{dmg}$ and $F_{th}(1)$, nano-voids could form during the first shot, requiring fewer number of pulses to reach the onset of ablation. The above conclusion is expected to apply qualitatively to \ch{WS2}. To understand quantitatively the difference of incubation between these two materials, a microscopic model involving multiple kinetic rate equations is needed.\cite{Emmert2010} While such a model better describes the physics of incubation, its implementation is challenging since many properties associated with these defects in 2D materials such as their energy levels and lifetimes are not known in literature or are speculative at best. The fast saturation and weak incubation establish \ch{MoS2} as an attractive material for high-throughput laser processing and strong field devices. 

\begin{acknowledgments}
This research was supported by the Ministry of Science and Technology (MOST) of Taiwan (Grant No's.\@ 109-2112-M-009-020-MY3, 109-2124-M-009-003-MY3, MOST-110-2119-M-A49-001-MBK) and the Center for Emergent Functional Matter Science (CEFMS) of NYCU supported by the Ministry of Education of Taiwan.
\end{acknowledgments}

\section*{Author Declarations}
The authors have no conflicts to disclose.

\section*{Data Availability Statement}
The data that supports the findings in this study are available within the article and its supporting information.

\bibliography{Multishot_Citations}

\clearpage
\includepdf[pages={1}]{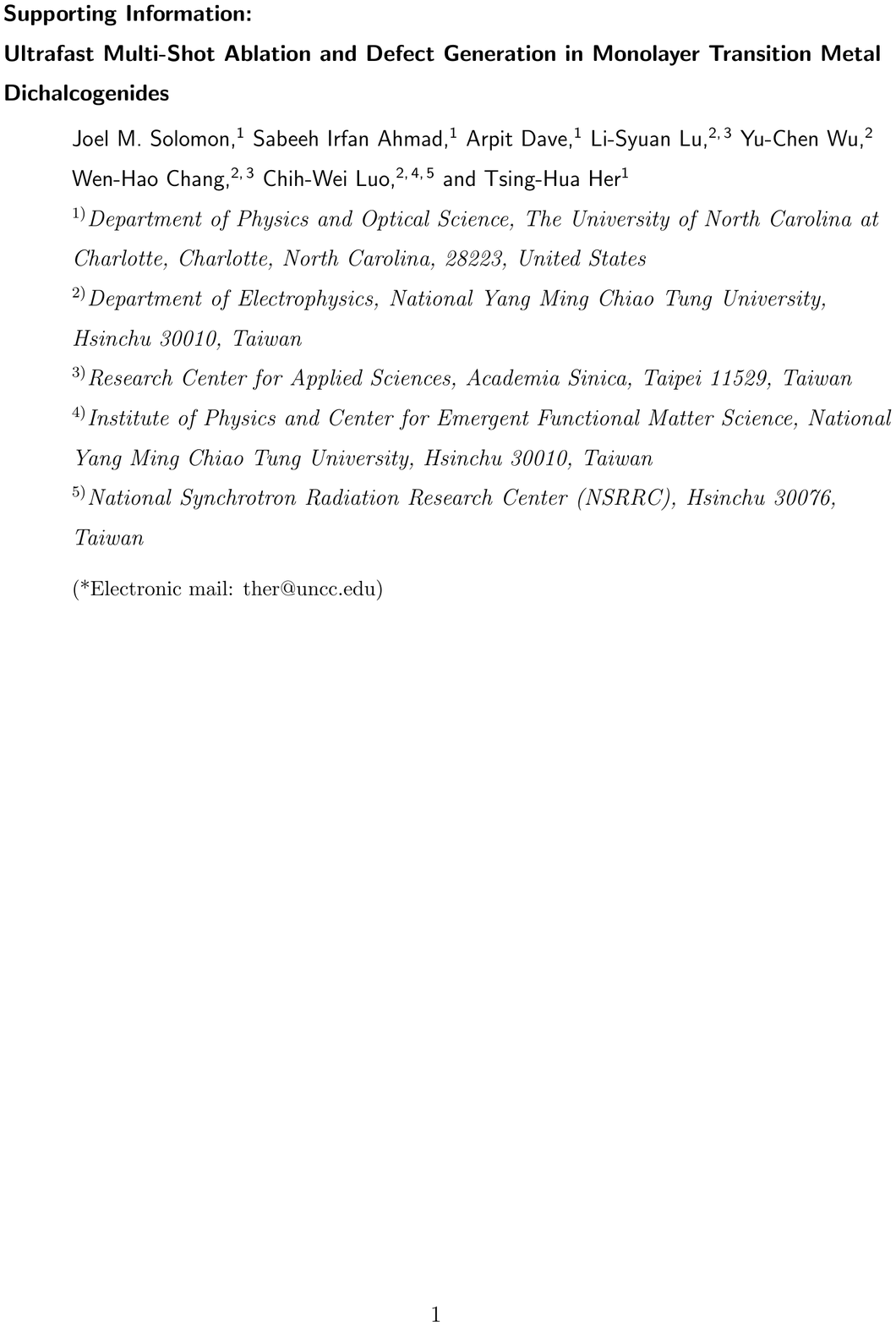}
\clearpage
\includepdf[pages={2}]{SI}
\clearpage
\includepdf[pages={3}]{SI}

\end{document}